\documentclass[12pt,dvips]{article}
\usepackage{epsfig}
\usepackage{longtable}
\usepackage{graphicx}
\usepackage{amsfonts}
\usepackage{amssymb}
\usepackage{amsthm}
\usepackage{amsmath}
\usepackage{cite} 

\usepackage[normalem]{ulem}

\usepackage{float}

\newcommand{\tev}{\,\, \mathrm{TeV}}
\newcommand{\gev}{\,\, \mathrm{GeV}}

\newcommand\be{\begin{equation}}
\newcommand\ee{\end{equation}}
\newcommand\bea{\begin{eqnarray}}
\newcommand\eea{\end{eqnarray}}
\newcommand\nn{\nonumber}

\newcommand{\ga}{\alpha}

\makeatletter
\def\lsim{\mathrel{\mathpalette\@versim<}}
\def\gsim{\mathrel{\mathpalette\@versim>}}
\def\@versim#1#2{\vcenter{\offinterlineskip
\ialign{$\m@th#1\hfil##\hfil$\crcr#2\crcr\sim\crcr } }}
\makeatother
\newcommand{\complex}{{{\rm I} \kern -.59em {\rm C}}}

\usepackage[active]{srcltx}
\usepackage[dvips]{color}

\begin{document}

\thispagestyle{empty}

\hfill CERN-PH-TH/2012-364
\vspace*{3cm}
\begin{center}
{\Large{\bf
Reduction of Couplings in the MSSM}}

\vspace*{1cm}
{\bf
M. Mondrag\'{o}n,$^{(1)}$
N.D.Tracas$^{(2)*}$ and G. Zoupanos$^{(2,3)*}$}
\end{center}
\vspace*{5mm}
\begin{center}
$^{(1)}$ Instituto de F\'isica, Universidad Nacional Aut\'{o}noma de M\'{e}xico,\\
Apdo. Postal 20-364,\\ Mexico 01000 D.F., M\'{e}xico\\
$^{(2)}$Theory Group, Physics Department, CERN,\\
1211 Geneva 23\\
$^{(3)}$ Max-Planck-Institut f\"{u}r Physik (Werner-Heisenberg-Institut)\\
F\"{o}hringer Ring 6 80805 M\"{u}nchen\\
and\\
Arnold-Sommerfeld-Center f\"{u}r Theoretische Physik\\
Department f\"{u}r Physik, Ludwig-Maximilians-Universit\"{a}t M\"{u}nchen
\end{center}

\vspace*{1cm}
\begin{center}
{\bf Abstract}
\end{center}

\noindent
In this paper, we first demonstrate the existence of renormalization group invariant relations among the top, bottom
Yukawa and the gauge colour couplings in the minimal supersymmetric SM.
Based on this observation and assuming furthermore the existence of a renormalization group invariant relation among
the trilinear couplings in the superpotential and the soft supersymmetry breaking sector,
we obtain predictions for the Higgs masses and the supersymmetric spectrum.

\noindent

\vfill
$^*$ On leave of absence from the Physics Department, National Technical University of Athens,
157 73 Zografou,  Athens, Greece.
\newpage

\section{Introduction}

With the recent discovery of the Higgs-like boson at the LHC
\cite{Aad:2012tfa++},
the new bounds on supersymmetric particles which place supersymmetry
at least at the TeV scale
\cite{Pravalorio:susy2012++},
and the new data on B physics \cite{Aaij:2012nna++}, the
search for theoretical scenarios beyond the Standard Model in which
all these experimental facts can be accomodated becomes more pressing.

Frameworks such as Superstrings and Noncommutative
Theories were developed aiming to provide a unified description
of all interactions, including gravity.  However, the main goal
from a unified description of interactions should be the
understanding of the present day free parameters of the Standard Model
(SM) in terms of a few fundamental ones, or in other words to achieve
{\it reduction of couplings} at a more fundamental
level. Unfortunately, the above theoretical frameworks have not provided
yet an understanding of the free parameters of the SM.

We have developed a complementary strategy in searching for a more
fundamental theory, possibly realized near the Planck scale, whose
basic ingredients are Grand Unified Theories (GUTs) and supersymmetry (SUSY) ,
but its consequences certainly go beyond the known ones
\cite{Kapetanakis:1992vx++,Mondragon:1993tw,Kubo:1996js}.
The method consists on searching for renormalization group invariant (RGI)
relations holding below the Planck scale, which in turn are preserved
down to the GUT scale.  An impressive aspect of the RGI relations is
that one can guarantee their validity to all-orders in perturbation
theory by studying the uniqueness of the resulting relations at
one-loop, as was proven in the early days of the programme of {\it
reduction of couplings}
\cite{Zimmermann:1984sx++}. Even more
remarkable is the fact that it is possible to find RGI relations among
couplings that guarantee finiteness to all-orders in perturbation
theory \cite{Lucchesi:1987he++}.  This
programme, called Gauge--Yukawa unification  (GYU) scheme, has been applied
to the dimensionless couplings of supersymmetric GUTs, such as gauge
and Yukawa couplings, with remarkable successes since it predicted
correctly the top quark and the Higgs masses in finite $N = 1$
supersymmetric SU(5) GUTs
\cite{Kapetanakis:1992vx++,Mondragon:1993tw,Kubo:1996js,Heinemeyer:2007tz++}.

Supersymmetry seems to be an essential feature of the GYU programme and
understanding its breaking becomes crucial, since the programme
has the ambition to supply the SM with predictions for several of its
free parameters. Indeed, the search for RGI relations was extended to
the soft supersymmetry breaking (SSB) sector of these theories
\cite{Kubo:1996js,Jack:1995gm}, which involves parameters of dimension
one and two.  Based conceptually and technically on the work of
ref.~\cite{Yamada:1994id}, considerable progress was made concerning
the renormalization properties of the SSB parameters
\cite{Hisano:1997ua,Jack:1997pa,Avdeev:1997vx++,Kazakov:1995cy,Kazakov:1997nf++,Kobayashi:1998jq}.
In ref.~\cite{Yamada:1994id} the powerful supergraph method
\cite{Delbourgo:1974jg++,Fujikawa:1974ay}
was applied to softly broken SUSY theories  using the ``spurion''
external space-time independent superfields
\cite{Girardello:1981wz,HelayelNeto:1984iv++}.

In the spurion method, a softly broken supersymmetric gauge theory is
considered as a supersymmetric one in which the various parameters
such as couplings and masses have been promoted to external
superfields that acquire ``vacuum expectation values''.
Thus, the $\beta$-functions of the parameters of the softly broken
theory are expressed in terms of partial differential operators
involving the dimensionless parameters of the unbroken theory.  By
transforming the partial differential operators involved into total
derivative operators it is possible to express all parameters in a RGI
way \cite{Kazakov:1997nf++,Kobayashi:1998jq}, and in
particular on the RGI surface which is defined
by the solution of the reduction equations.  Crucial to the success of
this programme is that the soft scalar masses
obey a sum rule \cite{Kawamura:1997cw,Kobayashi:1997qx}, which is RGI
to all orders in perturbation theory, both for the general GYU as for
the particular finite case \cite{Kobayashi:1998jq}.
Based on the above tools and results we would
like to apply the above programme in the case of MSSM.

\section{The Reduction of Couplings Method}

In this section we will briefly outline the reduction of couplings method. Any RGI
relation among couplings (i.e. which does not depend on the
renormalization scale $\mu$ explicitly) can be expressed, in the
implicit form $\Phi (g_1,\cdots,g_A) ~=~\mbox{const.}$, which has to
satisfy the partial differential equation (PDE)
\be
\frac{d \Phi}{dt}=
\sum_{a=1}^{A}\,\frac{\partial \Phi}{\partial g_{a}}
\frac{dg_a}{dt}=
\sum_{a=1}^{A} \,\frac{\partial \Phi}{\partial g_{a}}\beta_{a}=
{\vec \nabla} \Phi \cdot {\vec \beta} =0,
\ee
where $t=\ln\mu$ ($\mu$ being the renormalization scale) and $\beta_a$ is the $\beta$-function of $g_a$. This PDE is
equivalent to a set of ordinary differential equations, the so-called
reduction equations (REs)
\cite{Zimmermann:1984sx++,Oehme:1985jy},
\be \beta_{g} \,\frac{d
g_{a}}{d g} =\beta_{a}~,~a=1,\cdots,A~,
\label{redeq}
\ee
where $g$ and $\beta_{g}$ are the primary coupling and its
$\beta$-function, and the counting on $a$ does not include $g$. Since
maximally ($A-1$) independent RGI ``constraints'' in the
$A$-dimensional space of couplings can be imposed by the $\Phi_a$'s,
one could in principle express all the couplings in terms of a single
coupling $g$.  The strongest requirement is to demand power series
solutions to the REs,
\be g_{a} = \sum_{n=0}\rho_{a}^{(n)}\,g^{2n+1}~,
\label{powerser}
\ee
which formally preserve perturbative renormalizability. Remarkably,
the uniqueness of such power series solutions can be decided already
at the one-loop level
\cite{Zimmermann:1984sx++,Oehme:1985jy}. To illustrate
this, let us assume that the $\beta$-functions have the form
\bea
\beta_{a} &=&\frac{1}{16 \pi^2}[ \sum_{b,c,d\neq
  g}\beta^{(1)\,bcd}_{a}g_b g_c g_d+
\sum_{b\neq g}\beta^{(1)\,b}_{a}g_b g^2]+\cdots~,\nn\\
\beta_{g} &=&\frac{1}{16 \pi^2}\beta^{(1)}_{g}g^3+ \cdots~,
\eea
where
$\cdots$ stands for higher order terms, and $ \beta^{(1)\,bcd}_{a}$'s
are symmetric in $ b,c,d$.  We then assume that the $\rho_{a}^{(n)}$'s
with $n\leq r$ have been uniquely determined. To obtain
$\rho_{a}^{(r+1)}$'s, we insert the power series (\ref{powerser}) into
the REs (\ref{redeq}) and collect terms of $O(g^{2r+3})$ and find
\begin{equation}
\sum_{d\neq g}M(r)_{a}^{d}\,\rho_{d}^{(r+1)} =
\mbox{lower order quantities}~,
\end{equation}
where the r.h.s. is known by assumption, and
\begin{align}
M(r)_{a}^{d} &=3\sum_{b,c\neq g}\,\beta^{(1)\,bcd}_{a}\,\rho_{b}^{(1)}\,
\rho_{c}^{(1)}+\beta^{(1)\,d}_{a}
-(2r+1)\,\beta^{(1)}_{g}\,\delta_{a}^{d}~,\label{M}\\
0 &=\sum_{b,c,d\neq g}\,\beta^{(1)\,bcd}_{a}\,
\rho_{b}^{(1)}\,\rho_{c}^{(1)}\,\rho_{d}^{(1)}
+\sum_{d\neq g}\beta^{(1)\,d}_{a}\,\rho_{d}^{(1)}
-\beta^{(1)}_{g}\,\rho_{a}^{(1)}~.
\end{align}
 Therefore, the $\rho_{a}^{(n)}$'s for all $n > 1$ for a given set of $\rho_{a}^{(1)}$'s can be uniquely determined if
$\det M(n)_{a}^{d} \neq 0$  for all $n \geq 0$.

Our experience examining specific examples has taught us that the
various couplings in supersymmetric theories could have the same
asymptotic behaviour.  Therefore, searching for a power series solution
of the form (\ref{powerser}) to the REs (\ref{redeq}) is justified and
moreover, one can rely that keeping only the first terms a good
approximation is obtained in realistic applications.

\section{Sum Rule for Soft Breaking Terms}

The method of reducing the dimensionless couplings has been
extended\cite{Kubo:1996js,Jack:1995gm}, as we have discussed in the introduction, to the soft
supersymmetry breaking (SSB) dimensionful parameters of $N = 1$
supersymmetric theories.  In addition it was found \cite{Kawamura:1997cw,Kobayashi:1997qx} that
RGI SSB scalar masses in Gauge-Yukawa unified models satisfy a
universal sum rule.

Consider the superpotential given by
\be
W= \frac{1}{2}\,\mu^{ij} \,\Phi_{i}\,\Phi_{j}+
\frac{1}{6}\,C^{ijk} \,\Phi_{i}\,\Phi_{j}\,\Phi_{k}~,
\label{supot}
\ee
along with the Lagrangian for SSB terms
\be
-{\cal L}_{\rm SSB} =
\frac{1}{6} \,h^{ijk}\,\phi_i \phi_j \phi_k
+
\frac{1}{2} \,b^{ij}\,\phi_i \phi_j
+
\frac{1}{2} \,(m^2)^{j}_{i}\,\phi^{*\,i} \phi_j+
\frac{1}{2} \,M\,\lambda \lambda+\mbox{H.c.},
\ee
where the $\phi_i$ are the scalar parts of the chiral superfields $\Phi_i$, $\lambda$ are the gauginos
and $M$ their unified mass.

Let us recall that the one-loop $\beta$-function of the gauge coupling
$g$ is given by
\cite{Parkes:1984dh++}
  \bea \beta^{(1)}_{g}=\frac{d g}{d t} =
  \frac{g^3}{16\pi^2}\,[\,\sum_{i}\,T(R_{i})-3\,C_{2}(G)\,]~,
\label{betag}
\eea
where $C_{2}(G)$ is the quadratic Casimir of the adjoint representation of the associated
gauge group $G$. $T(R)$ is given by the relation $\textrm{Tr}[T^aT^b]=T(R)\delta^{ab}$ where
$T^a$ is the generators of the group in the appropriate representation.
Similarly the $\beta$-functions of $C_{ijk}$, by virtue of the non-renormalization theorem, are related to the
anomalous dimension matrix $\gamma^i_j$ of the chiral superfields as:
\be
\beta_C^{ijk} =
  \frac{d C_{ijk}}{d t}~=~C_{ijl}\,\gamma^{l}_{k}+
  C_{ikl}\,\gamma^{l}_{j}+
  C_{jkl}\,\gamma^{l}_{i}~.
\label{betay}
\ee
At one-loop level the anomalous dimension, $\gamma^{(1)}\,^i_j$ of the chiral superfield  is \cite{Parkes:1984dh++}
\be
\gamma^{(1)}\,^i_j=\frac{1}{32\pi^2}\,[\,
C^{ikl}\,C_{jkl}-2\,g^2\,C_{2}(R_{i})\delta_{ij}\,],
\label{gamay}
\ee
where $C_{2}(R_{i})$ is the quadratic Casimir of the
representation $R_{i}$, and $C^{ijk}=C_{ijk}^{*}$.
Then, the $N = 1$ non-renormalization theorem \cite{Wess:1973kz++,Fujikawa:1974ay} ensures
there are no extra mass and cubic-interaction-term renormalizations,
implying that the $\beta$-functions of $C_{ijk}$ can be expressed as
linear combinations of the anomalous dimensions $\gamma^i_j$.

Here we assume that the reduction equations admit power series solutions of the form
\be
C^{ijk} = g\,\sum_{n=0}\,\rho^{ijk}_{(n)} g^{2n}~.
\label{Yg}
\ee

In order to obtain higher-loop results instead of knowledge of
explicit $\beta$-functions, which anyway are known only up to
two-loops, relations among $\beta$-functions are required.

The progress made using the spurion technique, \cite{Delbourgo:1974jg++,Fujikawa:1974ay,Girardello:1981wz} leads to
the following  all-loop relations among SSB $\beta$-functions (in an
obvious notation),
\cite{Hisano:1997ua,Jack:1997pa,Avdeev:1997vx++,Kazakov:1997nf++}
\bea
\beta_M &=& 2{\cal O}\left(\frac{\beta_g}{g}\right)~,
\label{betaM}\\
\beta_h^{ijk}&=&\gamma^i{}_lh^{ljk}+\gamma^j{}_lh^{ilk}
+\gamma^k{}_lh^{ijl}\nn\\
&&-2\gamma_1^i{}_lC^{ljk}
-2\gamma_1^j{}_lC^{ilk}-2\gamma_1^k{}_lC^{ijl}~,\label{betah}\\
(\beta_{m^2})^i{}_j &=&\left[ \Delta
+ X \frac{\partial}{\partial g}\right]\gamma^i{}_j~,
\label{betam2}
\eea
where
\bea
{\cal O} &=&\left(Mg^2\frac{\partial}{\partial g^2}
-h^{lmn}\frac{\partial}{\partial C^{lmn}}\right)~,
\label{diffo}\\
\Delta &=& 2{\cal O}{\cal O}^* +2|M|^2 g^2\frac{\partial}
{\partial g^2} +\tilde{C}_{lmn}
\frac{\partial}{\partial C_{lmn}} +
\tilde{C}^{lmn}\frac{\partial}{\partial C^{lmn}}~,\\
(\gamma_1)^i{}_j&=&{\cal O}\gamma^i{}_j,\\
\tilde{C}^{ijk}&=&
(m^2)^i{}_lC^{ljk}+(m^2)^j{}_lC^{ilk}+(m^2)^k{}_lC^{ijl}~.
\label{tildeC}
\eea

The assumption, following \cite{Jack:1997pa}, that the relation among couplings
\be
h^{ijk} = -M (C^{ijk})'
\equiv -M \frac{d C^{ijk}(g)}{d \ln g}~,
\label{h2}
\ee
is RGI and furthermore, the use the all-loop gauge $\beta$-function of Novikov {\em et al.}\cite{Novikov:1983ee++} given by
\be
\beta_g^{\rm NSVZ} =
\frac{g^3}{16\pi^2}
\left[ \frac{\sum_l T(R_l)(1-\gamma_l /2)
-3 C_2(G)}{ 1-g^2C_2(G)/8\pi^2}\right]~,
\label{bnsvz}
\ee
lead to the all-loop RGI sum rule \cite{Kobayashi:1998jq} (assuming $(m^2)^i{}_j=m^2_j\delta^i_j$),
\begin{equation}
\begin{split}
m^2_i+m^2_j+m^2_k &=
|M|^2 \left\{~
\frac{1}{1-g^2 C_2(G)/(8\pi^2)}\frac{d \ln C^{ijk}}{d \ln g}
+\frac{1}{2}\frac{d^2 \ln C^{ijk}}{d (\ln g)^2}~\right\}\\
& +\sum_l
\frac{m^2_l T(R_l)}{C_2(G)-8\pi^2/g^2}
\frac{d \ln C^{ijk}}{d \ln g}~.
\label{sum2}
\end{split}
\end{equation}

Surprisingly enough, the all-loop result of Eq.(\ref{sum2}) coincides with
the superstring result for the finite case in a certain class of orbifold models \cite{Ibanez:1992hc++,Kobayashi:1997qx} if
\[
\frac{d \ln C^{ijk}}{d \ln g}=1~,
\]
as discussed in ref.~\cite{Mondragon:1993tw}.

\section{All-loop RGI Relations in the SSB Sector}

Lets us now see how the all-loop results on the SSB $\beta$-functions, Eqs.(\ref{betaM})-(\ref{tildeC}),
lead to all-loop RGI relations. We assume:\\
(a) the existence of a RGI surfaces on which $C = C(g)$, or equivalently that
\be
\label{Cbeta}
\frac{dC^{ijk}}{dg} = \frac{\beta^{ijk}_C}{\beta_g}
\ee
holds,  i.e. reduction of couplings is possible, and\\
(b) the existence of a RGI surface on which
\be
\label{h2NEW}
h^{ijk} = - M \frac{dC(g)^{ijk}}{d\ln g}
\ee
holds too in all-orders.\\
Then one can prove, \cite{Jack:1999aj,Kobayashi:1998iaa}, that the following relations are RGI to all-loops (note that in
both (a) and (b) assumptions above we do not rely on specific solutions of these equations)
\begin{align}
M &= M_0~\frac{\beta_g}{g} ,  \label{Mbeta} \\
h^{ijk}&=-M_0~\beta_C^{ijk},  \label{hbeta}  \\
b^{ij}&=-M_0~\beta_{\mu}^{ij},\label{bij}\\
(m^2)^i{}_j&= \frac{1}{2}~|M_0|^2~\mu\frac{d\gamma^i{}_j}{d\mu},
\label{scalmass}
\end{align}
where $M_0$ is an arbitrary reference mass scale to be specified shortly. The assumption that
\be
C_a\frac{\partial}{\partial C_a}
= C_a^*\frac{\partial}{\partial C_a^*} \label{dc/dc}
\ee
for a RGI surface $F(g,C^{ijk},C^{*ijk})$ leads to
\begin{equation}
\label{F}
\frac{d}{dg} = \left(\frac{\partial}{\partial g} + 2\frac{\partial}{\partial C}\,\frac{dC}{dg}\right)
= \left(\frac{\partial}{\partial g} + 2 \frac{\beta_C}{\beta_g}
\frac{\partial}{\partial C} \right)
\end{equation}
where Eq.(\ref{Cbeta}) has been used. Now let us consider the partial differential operator ${\cal O}$ in
Eq.(\ref{diffo}) which, assuming Eq.(\ref{h2}), becomes
\be
{\cal O} = \frac{1}{2}M\frac{d}{d\ln g}
\ee
In turn, $\beta_M$ given in Eq.(\ref{betaM}), becomes
\be
\beta_M = M\frac{d}{d\ln g} \big( \frac{\beta_g}{g}\big) ~, \label{betaM2}
\ee
which by integration provides us \cite{Karch:1998qa,Jack:1999aj} with the
generalized, i.e. including Yukawa couplings, all-loop RGI Hisano - Shifman relation \cite{Hisano:1997ua}
\be
 M = \frac{\beta_g}{g} M_0~, \label{M-M0}
\ee
where $M_0$ is the integration constant and can be associated to the
unification scale $M_U$ in GUTs or to the gravitino mass $m_{3/2}$ in
a supergravity framework. Therefore, Eq.(\ref{M-M0}) becomes the
all-loop RGI Eq.(\ref{Mbeta}).  Note that $\beta_M$ using
Eqs.(\ref{betaM2}) and (\ref{M-M0}) can be written as
\be \beta_M =
M_0\frac{d}{dt} (\beta _g/g)~.
\ee
Similarly
\be (\gamma_1)^i{}_j =
{\cal O} \gamma^i{}_j = \frac{1}{2}~M_0~\frac{d
  \gamma^i{}_j}{dt}~. \label{gammaO}
\ee
Next, from Eq.(\ref{h2}) and Eq.(\ref{M-M0}) we obtain
\be
 h^{ijk} = - M_0 ~\beta_C^{ijk}~,  \label{hm32}
\ee
while $\beta^{ijk}_h$, given in Eq.(\ref{betah}) and using Eq.(\ref{gammaO}), becomes \cite{Jack:1999aj}
\be
  \beta_h^{ijk} = - M_0~\frac{d}{dt} \beta_C^{ijk},
\ee
which shows that Eq.(\ref{hm32}) is all-loop RGI. In a similar way
Eq.(\ref{bij}) can be shown to be all-loop RGI.

Finally we would like to emphasize that under the same assumptions (a)
and (b) the sum rule given in Eq.(\ref{sum2}) has been proven
\cite{Kobayashi:1998jq} to be all-loop RGI, which (using Eq.(\ref{M-M0})) gives us
a generalization of Eq.(\ref{scalmass}) to be applied in
considerations of non-universal soft scalar masses, which are
necessary in many cases including the MSSM.

Having obtained the Eqs.(\ref{Mbeta})-(\ref{scalmass}) from
Eqs.(\ref{betaM})-(\ref{tildeC}) with the assumptions (a) and (b), we
would like to conclude the present section with some remarks.  First
it is worth noting the difference, say in first order in $g$, among
the possibilities to consider specific solution of the reduction
equations or just assume the existence of a RGI surface, which is a
weaker assumption. So in case we consider the reduction equation
(\ref{Cbeta}) without relying on a specific solution, the sum rule
(\ref{sum2}) reads
\be m^2_i+m^2_j+m^2_k = |M|^2 \frac{d\ln C^{ijk}}{d\ln g},
\label{sumrulenow}
\ee
and we find that
\be
\frac{d\ln C^{ijk}}{d\ln g} =\frac {g}{C^{ijk}}\frac{dC^{ijk}}{dg} =
\frac{g}{C^{ijk}}\frac{ \beta^{ijk}_C}{\beta_g},
\ee
which is clearly
model dependent. However assuming a specific power series solution of
the reduction equation, as in Eq.(\ref{powerser}), which in first
order in $g$ is just a linear relation among $C^{ijk}$ and $g$, we
obtain that
\be
\label{dcdg}
\frac{d\ln C^{ijk}}{d\ln g} = 1
\ee
and therefore the sum rule (\ref{sumrulenow}) becomes model
independent. We should also emphasize that in order to show
\cite{Jack:1997pa} that the relation
\be (m^2)^i{}_j =
\frac{1}{2}\frac{g^2}{\beta_g} |M|^2\frac{d\gamma^i{}_j}{dg},
\ee
which using Eq.(\ref{M-M0}) becomes Eq.(\ref{scalmass}), is RGI to
all-loops a specific solution of the reduction equations has to be
required. As it has already been pointed out above such a requirement
is not necessary in order to obtain the all-loop RG invariance of the
sum rule (\ref{sum2}).

As it was emphasized in ref \cite{Jack:1999aj} the set of the all-loop
RGI relations (\ref{Mbeta})-(\ref{scalmass}) is the one obtained in
the \textit{Anomaly Mediated SB Scenario} \cite{Randall:1998uk++}, by fixing
the $M_0$ to be $m_{3/2}$, which is the natural scale in the
supergravity framework.

A final remark concerns the resolution of the fatal problem of the
anomaly induced scenario in the supergravity framework, which  is
here solved thanks to the sum rule (\ref{sum2}), as it will become clear in
the next section.  Other solutions have been provided by introducing
Fayet-Iliopoulos terms \cite{Hodgson:2005en}.

\section{MSSM and RGI relations}

We would like now to apply the RGI relations to the SSB sector of the
MSSM, assuming power series solutions of the reduction equations at
the unfication scale. According to the analysis presented in Section 4
the RGI relations in the SSB sector hold, assuming the existence of RGI
surfaces where Eqs.(\ref{Cbeta}) and (\ref{h2NEW}) hold. We show first
 that Eq.(\ref{Cbeta}) indeed holds in the MSSM, then we assume
the validity of Eq.(\ref{h2NEW}) and examine the consequences in the
MSSM phenomenology.

Using a perturbative ansatz concerning the solutions of Eqs.(\ref{Cbeta}) and (\ref{h2NEW}), the set of
Eqs.(\ref{Mbeta})-(\ref{bij}) and Eq.(\ref{sumrulenow}) together with
Eq.(\ref{dcdg}), clearly hold.  Then one easily finds that
Eq.(\ref{h2NEW}) with (the first order) perturbative ansatz at the
unification scale leads to the condition
\begin{equation}
\label{eq:soft_Yukawa}
h^{ijk} =- M_U C^{ijk}, 
\end{equation}
where $M_U$ is the gaugino mass and $C^{ijk}$ are the Yukawa couplings, both
at the unification scale. Therefore, this assumption leads to Eqs.(\ref{eq:soft_Yukawa})
as boundary conditions at the unification scale.

In a similar way, starting from Eq.(\ref{bij}) and assuming that $\mu^{ij}$ are reduced in
favour of $g$, i.e.  that the reduction equation hold
\begin{equation}
\label{}
\beta_ {\mu}^{ij} = \beta_g d\mu^{ij}/dg  
\end{equation}
and moreover has power series type solutions, we obtain
\begin{equation}
\label{bij2}
b^{ij} = - M_U \mu^{ij}   
\end{equation}
as boundary conditions at the unification scale.

  Finally the sum rule (\ref{sumrulenow}) also holds  at the unification scale in the form,
\begin{equation}
\label{sumrulenow2}
m_i^2 + m_j^2 + m_k^2 = M_U^2  . 
\end{equation}
Therefore, the above Eqs.(\ref{eq:soft_Yukawa}),(\ref{bij2}) and (\ref{sumrulenow2})
have to be imposed as boundary conditions at the unification scale in the renormalization group equations
that govern the evolution of the SSB parameters.

Lets us now consider more specifically the MSSM, which is defined by the superpotential,
\begin{equation}
\label{supot2}
W = Y_tH_2Qt^c+Y_bH_1Qb^c+Y_\tau H_1L\tau^c+ \mu H_1H_2    ,    
\end{equation}
with soft breaking terms,
\begin{equation}
\label{}
\begin{split}
-\mathcal{L}_{SSB} &= \sum_\phi m^2_\phi\phi^*\phi+
\left[m^2_3H_1H_2+\sum_{i=1}^3 \frac 12 M_i\lambda_i\lambda_i +\textrm{h.c}\right]\\
&+\left[h_tH_2Qt^c+h_bH_1Qb^c+h_\tau H_1L\tau^c+\textrm{h.c.}\right] ,          
\end{split}
\end{equation}
where the last line refers to the scalar components of the corresponding superfield.
In general $Y_{t,b,\tau}$ and $h_{t,b,\tau}$ are $3\times 3$ matrices, but we work throughout in the
approximation that the  matrices are diagonal, and neglect the couplings
of the first two generations.

\noindent
{\it 5.1 Reduction of Couplings}\\
Assuming perturbative expansion of all three Yukawa couplings in favour of $\ga_3$ satisfying the
reduction equations
\begin{equation}
\label{tbRE}
\beta_{Y_{t,b,\tau}} = \beta_{g_3} \frac{dY_{t,b,\tau}}{d g_3} ,   
\end{equation}
we run into trouble since the coefficients of the $Y_\tau$ coupling turn imaginary. Therefore,
we take $Y_\tau$ at the GUT scale to be an independent variable. In that case, the coefficients of the expansions
(again at the GUT scale) 
\begin{align}
\frac{Y_t^2}{4\pi}&=c_1\frac{g_3^2}{4\pi}+c_2\left(\frac{g_3^2}{4\pi}\right)^2\label{alpha_t}\\
\frac{Y_b^2}{4\pi}&=p_1\frac{g_3^2}{4\pi}+p_2\left(\frac{g_3^2}{4\pi}\right)^2\label{alpha_b}
\end{align}
are given by
\be
\label{app_atau}
\begin{split}
c_1&=\frac{157}{175}+\frac 1{35}K_\tau=0.897 +0.029K_\tau
\\
p_1&=\frac{143}{175}-\frac{6}{35}K_\tau=0.817 -0.171K_\tau
\\
c_2&=\frac 1{4\pi}\,
\frac{1457.55 - 84.491 K_\tau - 9.66181 K_\tau^2 - 0.174927 K_\tau^3}
{818.943 - 89.2143 K_\tau - 2.14286 K_\tau^2}\\
p_2&=\frac 1{4\pi}\,
\frac{1402.52 - 223.777 K_\tau - 13.9475 K_\tau^2 - 0.174927 K_\tau^3}
{818.943 - 89.2143 K_\tau - 2.14286 K_\tau^2}
\end{split}
\ee
where
\be
K_\tau=Y_\tau^2/g_3^2
\label{ktau}
\ee
The important new observation is that the couplings $Y_t$,$Y_b$ and
$g_3$ are not only reduced, but  they provide predictions consistent
with the observed experimental values (as it will be explained later
in the discussion of Fig.(\ref{Mtopvsmbot})).

Given the above solutions of the reduction equations
\begin{equation}
\label{tbRE2}
\beta_{Y_{t,b}} = \beta_{g_3} \frac{dY_{t,b}}{d g_3} ,
\end{equation}
and assuming the validity of Eq.(\ref{h2NEW}) then, according to our earlier discussion, the following relations are
RGI
\begin{align}
M &= \frac{\beta_{g_3}}{g_3}  M_U ,
\label{eq:M_final}\\
h_{t,b} &= - M g_3 \frac{dY_{t,b}}{ dg_3},
\label{eq:h_final}\\
m_3^2 &= - M g_3 \frac{d\mu}{dg_3}  ,
\label{eq:m3_final}\\
m_i^2 + m_j^2 + m_k^2 &=  M^2,
\label{eq:sum_final}
\end{align}
where $i,j,k$ refer to the superfields appearing in the trilinear terms in the superpotential (\ref{supot2})\footnote{
There is another RGI term in the form of the b-parameter that could be included in Eq.(\ref{bij}) as was suggested in reference
\cite{Hodgson:2005en}.
This term would turn $m_3^2$ in Eqs.(\ref{eq:m3_final}) in a free parameter to be determined by the minimization of the
electroweak potential. Although we omit this term here, following other treatments in the literature, we plan to
include this possibility in a future
examination.}.

Note that in the application of the reduction of couplings in the MSSM that
we examine here, in the first stage we neglect the Yukawa couplings of the first two
generations, while we keep $Y_\tau$ and the gauge couplings $g_2$ and $g_1$, which cannot be reduced consistently,
as corrections. Therefore, strictly speaking, when we say above that
Eqs.(\ref{eq:M_final}-\ref{eq:sum_final}) are RGI we refer to the case that not only the first two
generations but also the $Y_\tau$, $g_2$ and $g_1$ are switched off.

In turn, since all gauge couplings in the MSSM meet at the unification
point, we are led to the following boundary conditions at the
GUT scale:
\begin{align}
Y_t^2&=c_1 g_U^2 + c_2 g_U^4/(4\pi) \quad\textrm{and}\quad
Y_b^2=p_1 g_U^2 + p_2 g_U^4/(4\pi)
\label{Y_t_Y_b} \\
h_{t,b} &= - M_U Y_{t,b}   , \label{eq:htb-MU}
\\
m_3^2 &= - M_U \mu          ,\label{m32}
\end{align}
where $c_{1,2}$ and $p_{1,2}$ are the solutions of the algebraic
system of the two reduction equations (\ref{tbRE}) taken at the GUT
scale (while keeping only the first term\footnote{ The second term can
  be determined once the first term is known.} of the perturbative
expansion of the Yukawas in favour of $g_3$ for
Eqs.(\ref{eq:htb-MU}) and (\ref{m32})), and a set of
equations resulting from the application of the sum rule
(\ref{sumrulenow2})
\begin{align}
m_{H_2}^2 + m_Q^2 + m_{t^c}^2 = M_U^2,\label{sum_rule_1}
\\
m_{H_1}^2 + m_Q^2 + m_{b^c}^2 = M_U^2 ,\label{sum_rule_2}
\end{align}
noting that the sum rule introduces four free parameters.

\begin{figure}[H]
\vspace{0.7cm}
\centering
\includegraphics[scale=0.4]{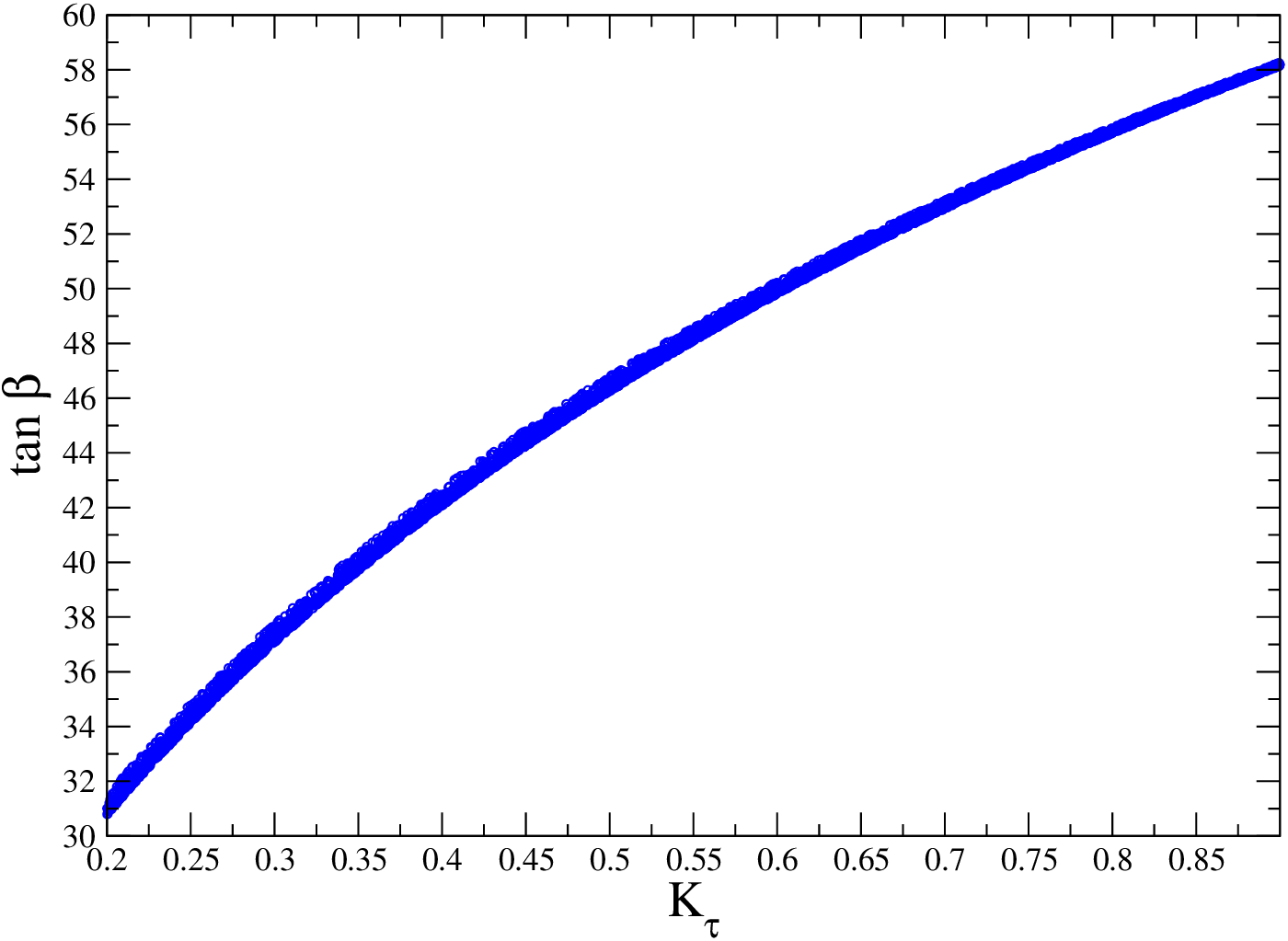}
\caption{Required values of $\tan\beta$ as a function of
  $K_\tau=Y_\tau^2/g_3^2$ in order to get the experimentally accepted tau mass.}
\label{aktauvstanbeta}
\end{figure}

\section{Discussion and Conclusions}
In the present paper we have made a new important observation, that
the $Y_t$, $Y_b$ and $\ga_3$ obey RGI relations within the
MSSM. Therefore, they can be reduced and can be considered as
parameters dependent among themselves.  This ``reduced'' system holds
at all scales, and thus serve as boundary conditions of the RGEs of
the MSSM at the unification scale, where we assume that the gauge
couplings meet.  With these boundary conditions we run the MSSM RGEs
down to the SUSY scale, which we take to be the geometrical average of
the stop masses, and then run the SM RGEs down to the electroweak
scale ($M_Z$), where we compare with the experimental values of the
third generation quark masses.  The RGEs are taken at two-loops for
the gauge and Yukawa couplings and at one-loop for the soft breaking
parameters.  We let $M_U$ and $|\mu|$ at the unification scale to vary
between $\sim 1 \tev \sim 11 \tev$, for the two possible signs of
$\mu$. In evaluating the $\tau$ and bottom masses we have taken into
account the one-loop radiative corrections that come from the SUSY
breaking \cite{ Carena:1994bv++}.  These corrections have a dependence on the soft
breaking parameters, in particular for large $\tan\beta$ they can give
sizeable contributions to the bottom quark mass.

\begin{figure}[H]
\centering
\vspace{0.7cm}
\includegraphics[scale=0.4]{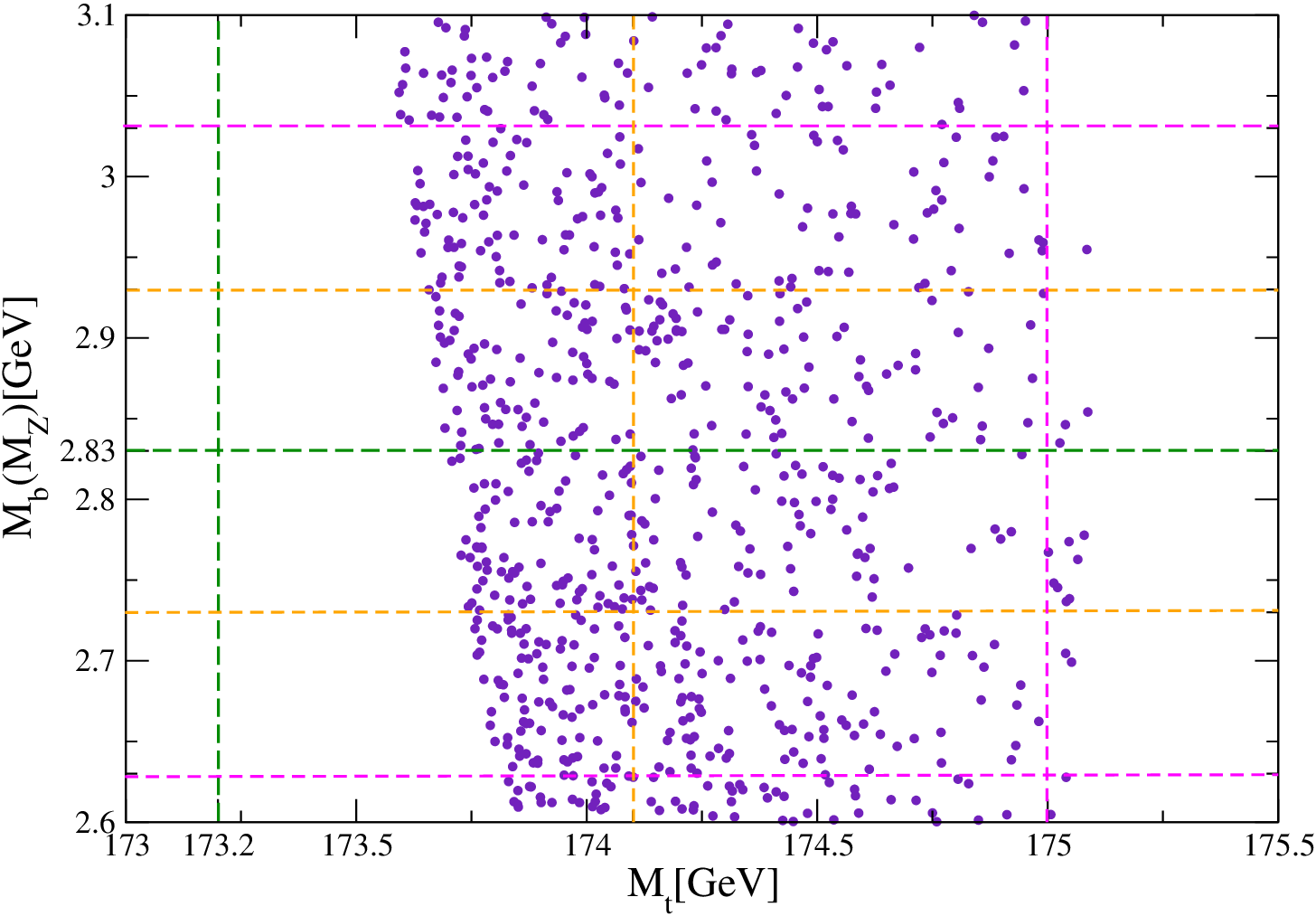}
\caption{Using the regions of values for $K_\tau=Y_\tau^2/g_3^2$ and $\tan\beta$ which give experimentally accepted tau mass, this figure shows
the resulted points in the $(M_t,M_b)$ phase space. The central value
(green dashed lines), as well as the 1 and 2$\sigma$ deviation (orange
and magenta lines respectively), for the top and bottom masses is
also  drawn.}
\label{Mtopvsmbot}
\end{figure}

The observation that $Y_t$, $Y_b$ and $\ga_3$  are a reduced system is best demonstrated in
Fig.(\ref{Mtopvsmbot}), where we plot the predictions for the top
quark mass, $M_{t}$, and the bottom quark mass, $M_{b}$, as they result from
Eqs.(\ref{alpha_t}) and (\ref{alpha_b}) with $c_{1,2}$ and $p{_1,_2}$
given in Eq.(\ref{app_atau}), for $\textrm{sign}(\mu)=-$. As one can see the predicted values
agree comfortably with the corresponding experimental values within 1$\sigma$.
Recall that
$Y_{\tau}$ is not reduced  and is a free parameter in this analysis.
In Fig.~(\ref{aktauvstanbeta}) we present a plot  relating the values of
$\tan\beta$ and $K_\tau=Y_\tau^2/g_3^2$
which are compatible with the observed experimental value
of the tau mass $M_{\tau}$ (fixed at its experimental central value).
In  the case that  $\textrm{sign}(\mu)=+$,  there is no value for $K_\tau$ where both the top and the bottom
quark masses agree simultaneously with their experimental value,
therefore we only consider the negative sign of $\mu$ from now on.

The parameter $K_\tau$ is further constrained by allowing only the
values that are also compatible  with the top and bottom quark masses
within 1 and $2\sigma$ of their central experimental value.
We use the experimental value of the top quark pole mass as
\cite{Group:2009ad++}
\be
M_t^{\rm exp} = (173.2 \pm 0.9) \gev ~.
\ee
The bottom mass is calculated at $M_Z$ to avoid uncertainties that
come from running down to the pole mass and, as previously mentioned,
the SUSY radiative corrections both to the tau and the bottom quark masses
have been taken into account \cite{Nakamura:2010zzi}
\be M_b(M_Z) =
(2.83 \pm 0.10) \gev .  \ee

\begin{figure}[H]
\centering
\vspace{0.7cm}
\includegraphics[scale=0.4]{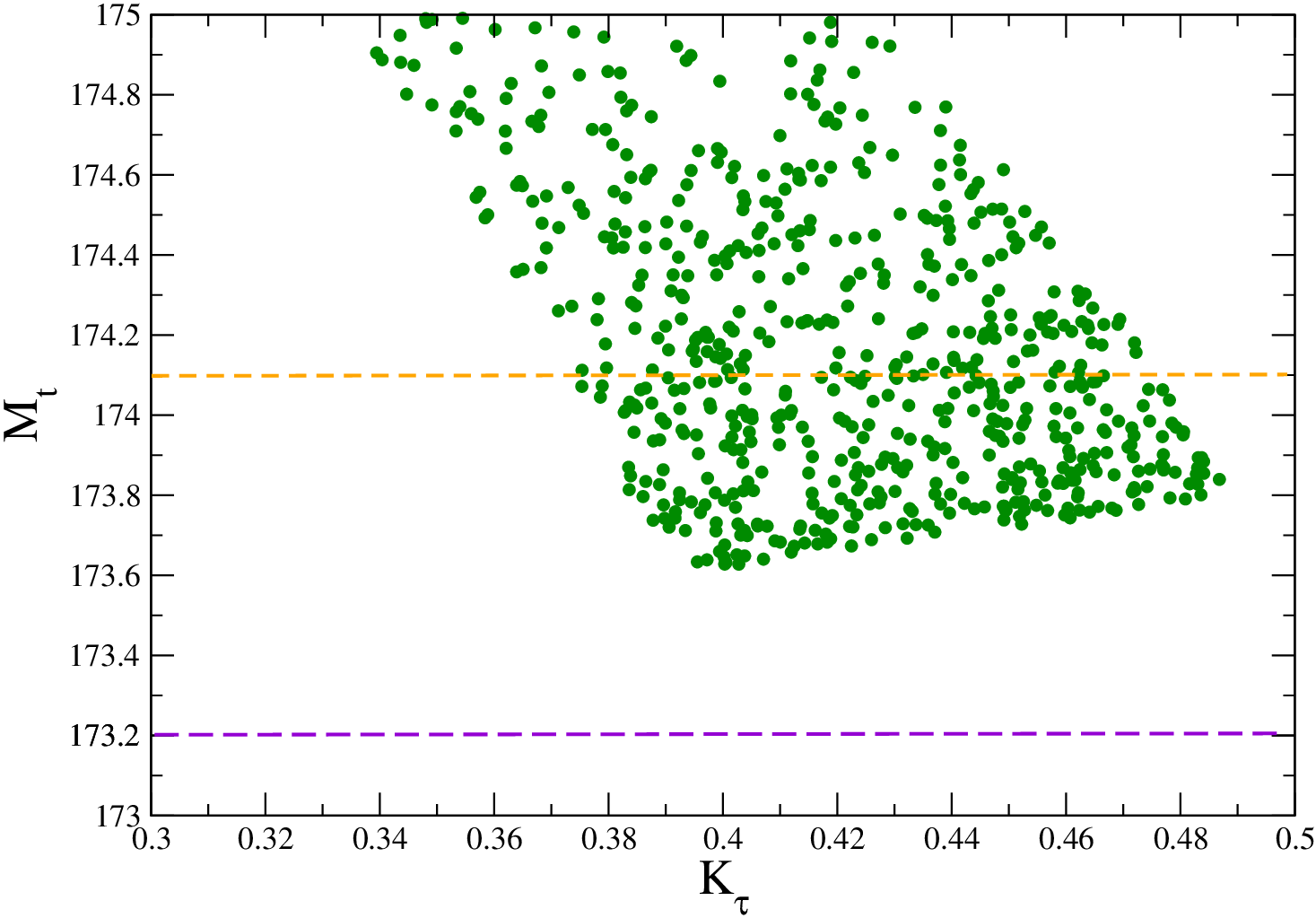}
\caption{The top mass as a function of $K_\tau=Y_\tau^2/g_3^2$, the
  purple dashed line is the experimental central value and the orange
  one is the $1\sigma$ value.}
\label{constrain-AKTAUvsMTOP}
\end{figure}

In Fig.(\ref{constrain-AKTAUvsMTOP}), we show these constrained
$K_\tau$ values plotted against $M_{t}$ (its central value corresponds
to the purple dashed line), within 1$\sigma$ (orange dashed lines), and
2$\sigma$ (upper border of the graph), where also $M_b$ is constrained to
be within 1 and $2\sigma$ of its experimental value. We can do the
same for $M_{b}$ but we prefer to present in Fig.(\ref{Mtopvsmbot})
the values of $M_{t}$ vs $M_{b}$ for the constrained $K_\tau$ values.
From Fig.~(\ref{Mtopvsmbot}) it can be clearly seen that
there is a set of values for the parameter $K_\tau$ where both $M_{t}$
and $M_{b}$ agree simultaneously within 1$\sigma$ of their
experimental values, for the boundary conditions given by the reduced
system $Y_t$, $Y_b$ and $\ga_3$.

\begin{figure}[H]
\centering
\vspace{0.7cm}
\includegraphics[scale=0.4]{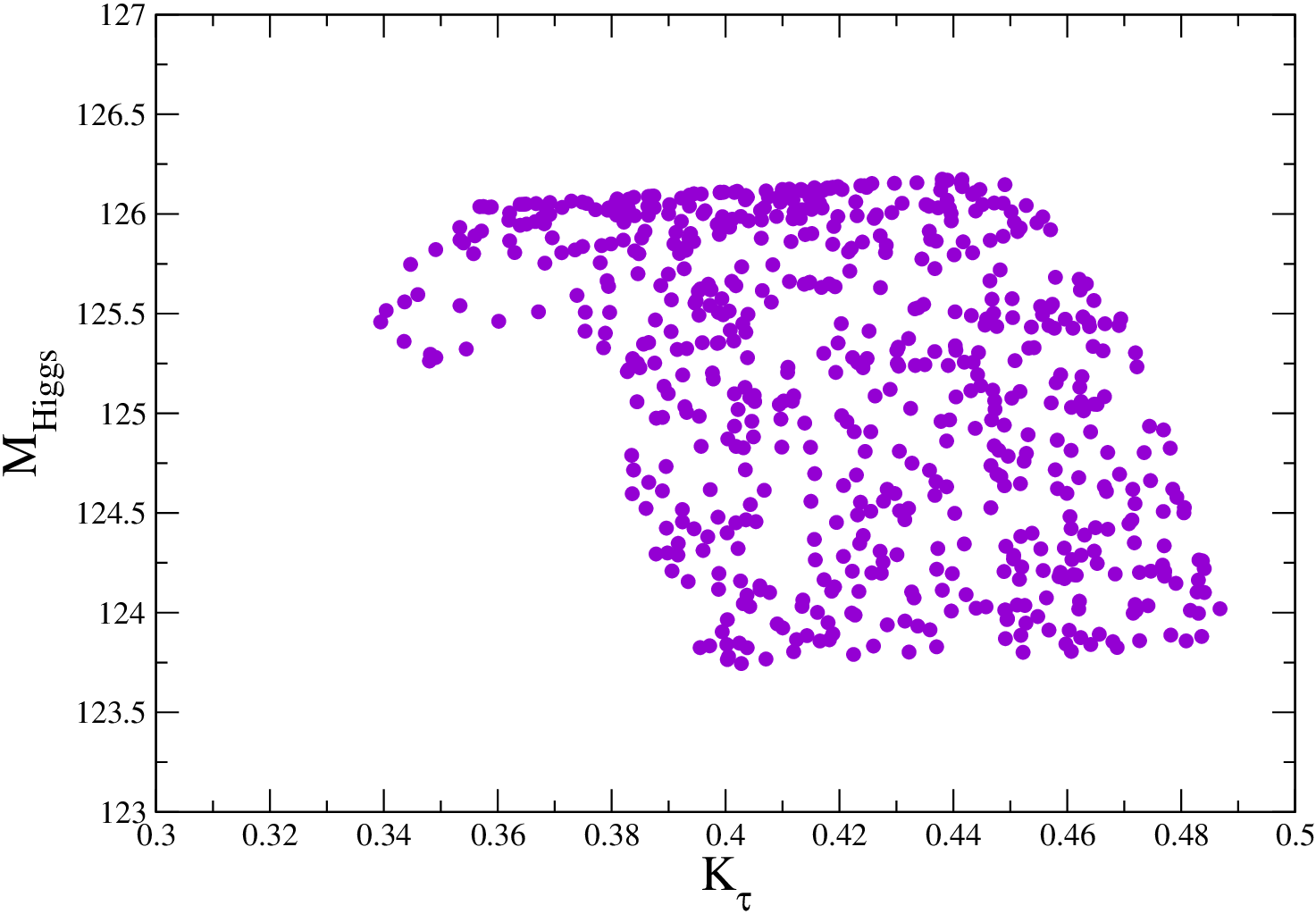}
\caption{The Higgs mass as a function of $K_\tau=Y_\tau^2/g_3^2$}
\label{constrain-AKTAUvsMHiggs}
\end{figure}

\begin{figure}[!b]
\centering
\includegraphics[scale=0.4]{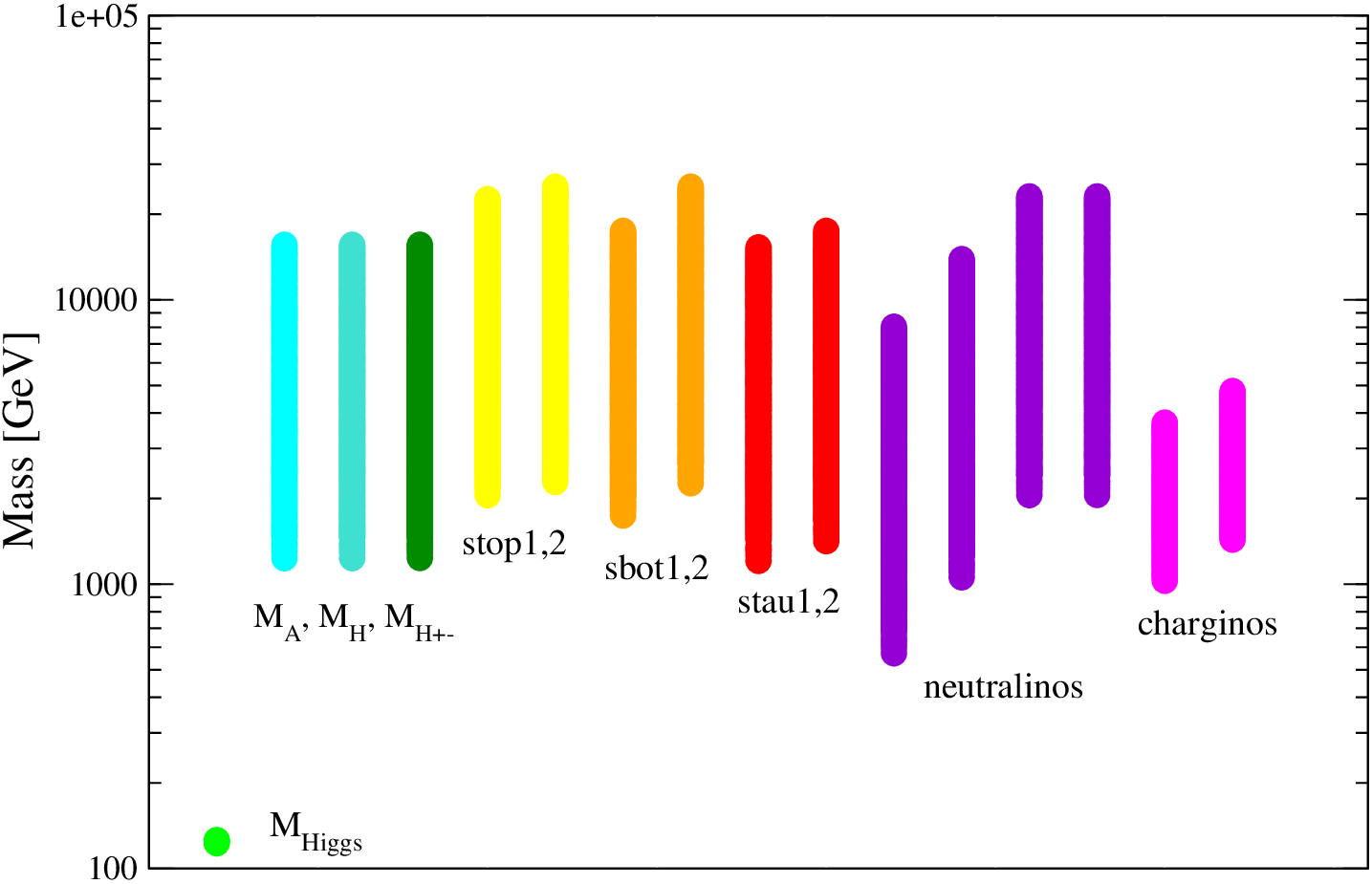}
\caption{The Higgs mass and s-spectrum for values of $M_U ~\sim 1.3
  \tev$ to $\sim 11 \tev$.}
\label{spectrum-b2}
\end{figure}

Finally, assuming the validity of Eq.(\ref{h2NEW}) for the corresponding couplings
to those that have been reduced before, we calculate
the Higgs mass as well as the whole Higgs and sparticle spectrum using
Eqs.(\ref{Y_t_Y_b})-(\ref{sum_rule_2}),
and we
present them in Figs.(\ref{constrain-AKTAUvsMHiggs}) and
(\ref{spectrum-b2}).
The Higgs mass was calculated using a ``mixed-scale'' one-loop RG approach,
which is known to be a very good approximation to the full
diagrammatic calculation \cite{Carena:2000dp++}.

From Fig.~(\ref{constrain-AKTAUvsMHiggs}) we notice that the lightest
Higgs mass is in the range $123.7$ - $126.3$~GeV, where the
uncertainty is due to the variation of $K_\tau$, the gaugino mass
$M_U$ and the variation of the scalar soft masses, which are however
constrained by the sum rules (\ref{sum_rule_1}) and
(\ref{sum_rule_2}).  The gaugino mass $M_U$ is in the range $\sim
1.3\tev \sim 11 \tev$, the lower values having been discarded since
they do not allow for radiative electroweak symmetry breaking.  The
variation of $K_\tau$ is in the range $\sim 0.37 \sim 0.49$ in order
to agree with the experimental values of the bottom and top masses at
1$\sigma$, and $\sim 0.34 \sim 0.49$ if the agreement is at the
2$\sigma$ level. To the lightest Higgs mass value one has to add at least $\pm 2$~GeV
coming from unknown higher order corrections \cite{Degrassi:2002fi}.
Therefore it is in excellent agreement with the experimental results
of ATLAS and CMS \cite{Aad:2012tfa++}.

From Fig.(\ref{spectrum-b2}) we find that the masses of the heavier
Higgses have relatively high values, above the TeV scale. In addition
we find a generally heavy supersymmetric spectrum starting with a
neutralino as LSP at $\sim 500$ GeV and comfortable agreement with the
LHC bounds due to the non-observation of  coloured supersymmetric particles~\cite{Pravalorio:susy2012++}.
Finally note that although the $\mu < 0$ found in our analysis would disfavour
the model in connection with the anomalous magnetic moment of the muon, such
a heavy spectrum gives only a negligible correction to the SM prediction.
We plan to extend our analysis by examining the restrictions that will be
imposed in the spectrum by the B-physics and CDM constraints.

\vspace*{1cm}\noindent \textit{\textbf{Acknowledgements.}}

One of us (G.Z.) would like to thank for discussions W. Hollik,
S. Heinemeyer, D. Luest, C. Mu\~noz, G. Ross, K. Sibold and for
discussions
and strong encouragement from R. Stora and W. Zimmermann.\\
The work of N.D.T. and G.Z. is supported by the European Union's ITN
programme UNILHC. The work of G.Z. ia also supported by the
Research Funding Program ARISTEIA, Higher Order
Calculations and Tools for High Energy Colliders, HOCTools
(co-financed by the European Union (European Social Fund ESF) and
Greek national funds through the Operational Program Education and
Lifelong Learning of the National Strategic Reference Framework
(NSRF)).  The work of M.M. is supported by mexican grants PAPIIT
IN113712 and Conacyt 132059.

\end{document}